\documentclass[aps,prb,twocolumn,superscriptaddress,showpacs]{revtex4}
\usepackage{graphics}
\usepackage{graphicx}
\usepackage[english]{babel}
\usepackage{color}

\begin{document}

\title{Effect of quenched disorder in the
entropy-jump at the first-order vortex phase transition of
Bi$_{2}$Sr$_{2}$CaCu$_{2}$O$_{8 + \delta}$ }

\author{M. I. Dolz}
\affiliation{Universidad Nacional de San Luis, San Luis,
Argentina}

\author{P. Pedrazzini}
\affiliation{Low Temperature Division, Centro At\'{o}mico
Bariloche, CNEA, Argentina}

\author{H. Pastoriza}
\affiliation{Low Temperature Division, Centro At\'{o}mico
Bariloche, CNEA, Argentina}

\author{M. Konczykowski}
\affiliation{Laboratoire des Solides Irradi\'{e}s, Ecole
Polytechnique, Palaiseau, France}

\author{Y. Fasano}
\affiliation{Low Temperature Division, Centro At\'{o}mico
Bariloche, CNEA, Argentina}

\date{\today}

%\keywords{vortex matter, mesoscopic physics, layered superconductors}

\begin{abstract}

We study the effect of quenched disorder in the thermodynamic
magnitudes entailed in the first-order vortex phase transition of
the extremely layered Bi$_{2}$Sr$_{2}$CaCu$_{2}$O$_{8 + \delta}$
compound. We track the temperature-evolution of the enthalpy and
the entropy-jump at the vortex solidification transition by means
of AC local magnetic measurements. Quenched
disorder is introduced to the pristine samples by means of
heavy-ion irradiation with Pb and Xe producing a random
columnar-track pins distribution with different densities
(matching field $B_{\Phi}$). In contrast with previous
magneto-optical reports, we find that the first-order phase
transition persists for samples with $B_{\Phi}$ up to 100\,Gauss.
For very low densities of quenched disorder (pristine samples),
the evolution of the thermodynamic properties can be
satisfactorily explained considering a negligible effect of
pinning and only electromagnetic coupling between pancake vortices
lying in adjacent CuO planes. This
description is not satisfactory on increasing
magnitude of quenched disorder.

%PACS numbers: 74.25.Uv, 74.25.Ha
\end{abstract}

\maketitle
\section{Introduction}

The vortex solidification or first-order phase transition (FOT)
\cite{Pastoriza94a,Zeldov95a} dominates the magnetic phase diagram
of pristine samples of anisotropic high-temperature
superconductors. The location of the FOT line and the
temperature-evolution of the thermodynamic magnitudes at the
transition are given by the balance between thermal,
vortex-quenched disorder and inter-vortex interaction energies.
All these terms are strongly dependent on the anisotropy parameter
of the material that determines the degree of layerness of vortex
matter. In the case of the extremely-anisotropic
Bi$_{2}$Sr$_{2}$CaCu$_{2}$O$_{8}$ compound, when applying a field
along the  $c$-axis individual vortices are composed of  a stack
of pancake vortices lying in CuO planes. Within one vortex line,
pancake vortices couple between layers due to electromagnetic and
Josephson interactions \cite{Blatter}. For temperatures $T<T_{\rm
FOT}$ the  stable phase in this compound is a vortex solid with
quasi long-range transversal positional order
\cite{Fasano1999,Fasano2008} and long-range coupling between the
pancake vortices concatenating a flux quantum. At the transition
temperature $T_{\rm FOT}$,  a decoupling process of pancake
vortices from adjacent layers takes place within the same stack
\cite{Colson2003}. There is still some controversy as to whether
the high-temperature phase is a liquid \cite{Nelson1988} or a
decoupled gas \cite{Glazman1991} of pancake vortices with reduced
shear viscosity \cite{Pastoriza1995}.

Nevertheless, the FOT  is a single-vortex transition in which the
relative importance of the two types of interactions between
pancake vortices determine the location of the $T_{\rm FOT}$ line
\cite{Blatter}. Previous works on pristine samples found a sudden
decrease of enthalpy and a divergence of the entropy for $T_{\rm
FOT} \sim T_{\rm c}$ by applying DC and AC magnetometry
\cite{Zeldov95a,Morozov1996}. These results were theoretically
interpreted as due to the Josephson interlaying coupling playing a
determinant role when the temperature of the transition approaches
$T_{\rm c}$ \cite{Dodgson1998}. This theoretical study does not
take into account the effect of the vortex-quenched disorder
interaction (pinning) that plays a relevant role in most cases.

Indeed, the magnitude of quenched disorder is a control parameter
that significantly affects the location of the FOT line. A largely
used method to introduce disorder with different magnitude is the
irradiation of samples with heavy ions resulting in the creation
of columnar tracks or defects (CD) \cite{Civale1997}. This method
generates a random distribution of strong pinning centers with
a density that can be adjusted by the irradiation dose. The
density of CDs is normally expressed in units of matching fields,
$B_{\Phi}=\Phi_{0}/a_{\rm CD}^{2}$, with $a_{\rm CD}$ the average spacing between
tracks. Heavy-ion-irradiated Bi$_{2}$Sr$_{2}$CaCu$_{2}$O$_{8}$
samples with a low density of CDs present a solid vortex phase
that spans a larger temperature-region on increasing the magnitude
of quenched disorder \cite{Banerjee2003,Menghini2003}.  Previous
magneto-optical measurements indicate the transition to the
high-temperature phase is no longer of first-order for matching
fields of roughly 100\,Gauss \cite{Banerjee2003,Menghini2003}. In
the latter case, a continuous change in the sample local induction
is observed on crossing the vortex solidification line, in
contrast to the sudden jump in $B$ detected for pristine
samples.

\begin{figure*}[ttt]
\includegraphics[width=0.95\textwidth]{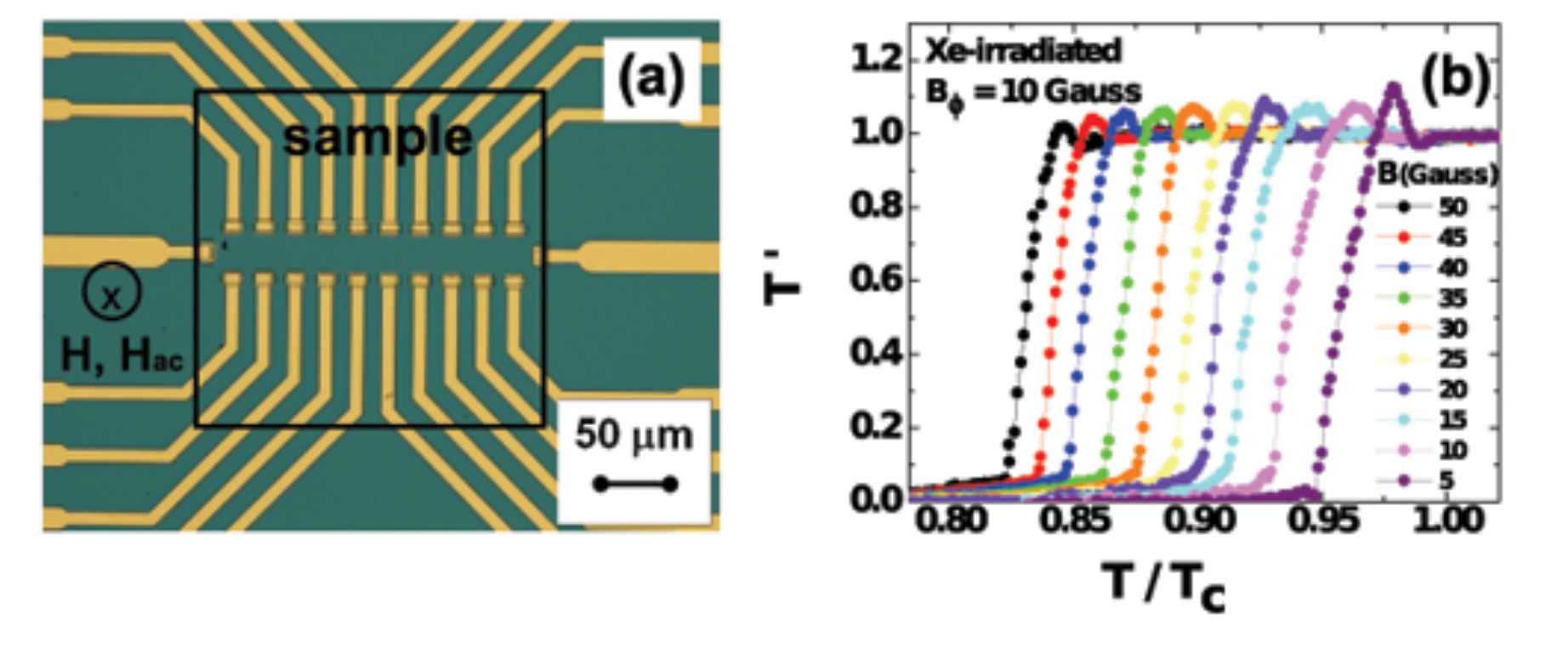}
\caption{ (a) Schematics of the experimental configuration:
picture of one of the 11-probe Hall arrays used for the AC
magnetometry measurements with a in-scale representation of the
sample size and location. The direction of the DC and AC fields
with respect to the sample and Hall probe are indicated. (b)
Temperature-evolution of the transmittivity for different applied
fields in the case of the Xe-irradiated
Bi$_{2}$Sr$_{2}$CaCu$_{2}$O$_{8}$ sample. The AC ripple field used
for this measurements had a magnitude of 2\,Gauss and a frequency
of 7\,Hz. (color figure online). \label{figure1}}
\end{figure*}

In this work we study the main thermodynamic magnitudes entailed
in the vortex solidification transition of
Bi$_{2}$Sr$_{2}$CaCu$_{2}$O$_{8}$ by means of local AC
magnetometry. We investigate the nature of the transition on
increasing the density of quenched disorder in one order of
magnitude, $B_{\rm \Phi}=10$ and 100\,Gauss for samples irradiated
with Xe and Pb ions. We focus on the effect of the magnitude of
disorder (directly proportional to the density of CDs) on the
entropy-jump, $\Delta S$, and on the enthalpy
of the transition, proportional to $\Delta B$.

\section{Experimental}

The nearly optimally-doped Bi$_{2}$Sr$_{2}$CaCu$_{2}$O$_{8}$
single-crystals used in this work were grown by means of the
traveling-solvent floating zone technique as discused in
Ref.\,\cite{Li94a}. The samples were irradiated with Pb and Xe
heavy-ions at Ganil, France. In this study we measure a pristine
sample, a Xe-irradiated sample with $B_{\Phi}=10$\,Gauss, and a
Pb-irradiated sample with $B_{\Phi}=100$\,Gauss. The samples were
mounted onto 2D-electron-gas Hall-sensor arrays
photolithographically fabricated from GaAs/AlGaAs
heterostructures. Each sensor has an active area of $6
\times 6$\,$\mu$m$^{2}$ and samples have typical areas of $200
\times 200$\,$\mu$m$^2$ and thickness of 30-50\,$\mu$m. DC and
AC-ripple magnetic fields, $H$ and $H_{\rm ac}$, are applied
parallel to the $c$-axis of the samples, namely perpendicular to
the Hall probe surface. The AC ripple fields have magnitudes of
1-2\,Gauss and frequencies up to 15\,Hz. A schematic
representation of the measurement configuration for one of the
probes used is shown in Fig.\,\ref{figure1} (a).

The local magnetic response was characterized by measuring the
first harmonic of the AC induction by means of a
digital-signal-processing lock-in technique \cite{Dolz2014}. We
obtained the transmittivity by normalizing the in-phase
component of the first harmonic signal as $T'=[B'(T) - B'(T \ll
T_{\rm c})]/[B'(T>T_{\rm c}) - B'(T \ll T_{\rm c})]$
\cite{Gilchrist1993}. This magnitude is extremely sensitive to
discontinuities in the local induction associated to first-order
magnetic transitions. In order to track the $H\,-\,T$ location of
the solidification transition in Bi$_{2}$Sr$_{2}$CaCu$_{2}$O$_{8}$
vortex matter we measured $T'$ as a function of temperature for
various DC applied fields.

\section{Results and discussion}

Figure \ref{figure1} (b) shows a set of transmittivity data for
the Xe-irradiated sample with $B_{\Phi}=10$\,Gauss and DC fields
ranging from 5 to 50\,Gauss. The so-called paramagnetic peak observed
in the curves is considered as the fingerprint of the first-order
vortex solidification transition since it develops at the same
$T_{\rm FOT}$ as the jump in $B$ detected in DC hysteresis loops
\cite{Morozov1996}. In addition, the temperature-location of this
paramagnetic peak is frequency-independent. The $T_{\rm FOT}$
temperature, taken at the maximum of the peak, shifts towards
lower temperatures on increasing field. The peaks are sharp and
have an amplitude that slightly decreases on increasing field. The
same field and temperature evolution was observed for pristine
as well as the Pb and Xe-irradiated samples.

\begin{figure*}[ttt]
\includegraphics[width=0.95\textwidth]{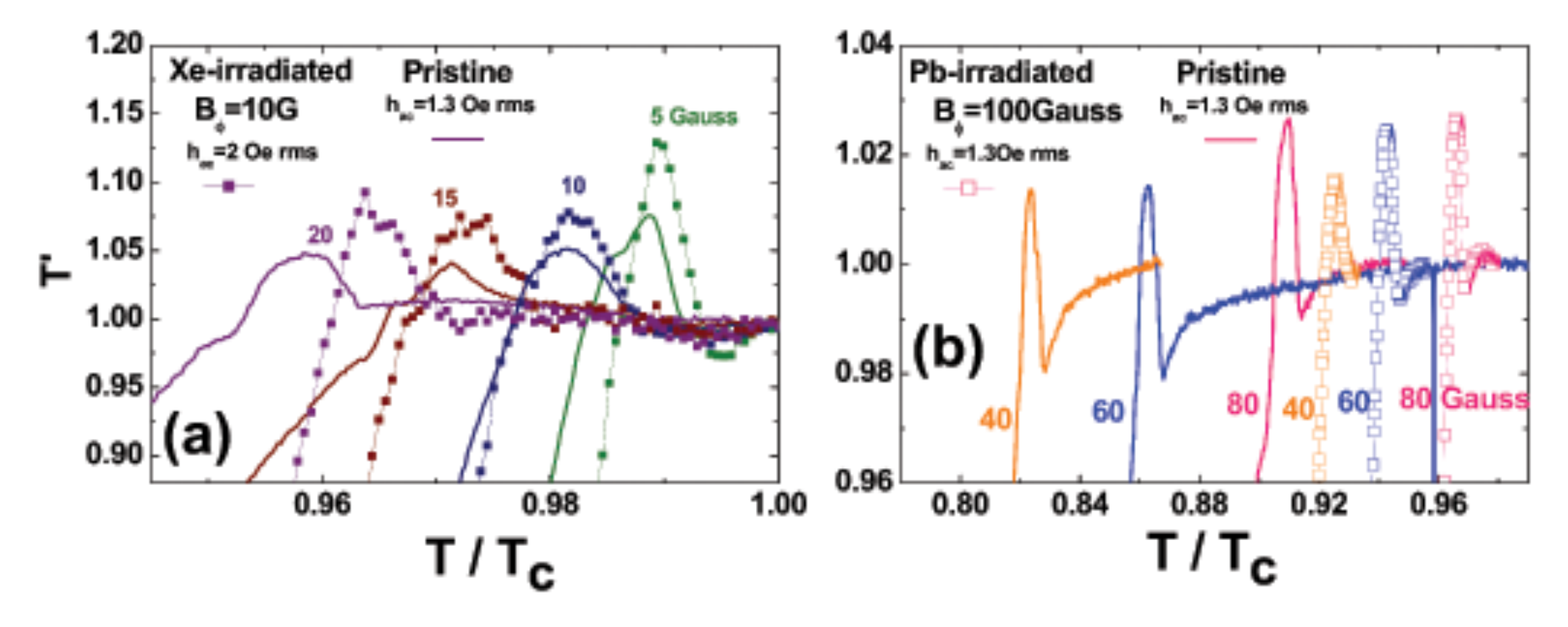}
\caption{Temperature-evolution of the transmittivity for low
fields for irradiated in comparison to pristine
Bi$_{2}$Sr$_{2}$CaCu$_{2}$O$_{8}$ samples: (a) Xe-irradiated
($B_{\Phi}=10$\,Gauss) and (b) Pb-irradiated
($B_{\Phi}=100$\,Gauss) samples.
The amplitudes of the AC ripple fields are as indicated and the
frequency is of 7\,Hz. (color figure online). \label{figure2}}
\end{figure*}

\begin{figure*}[ttt]
\includegraphics[width=0.95\textwidth]{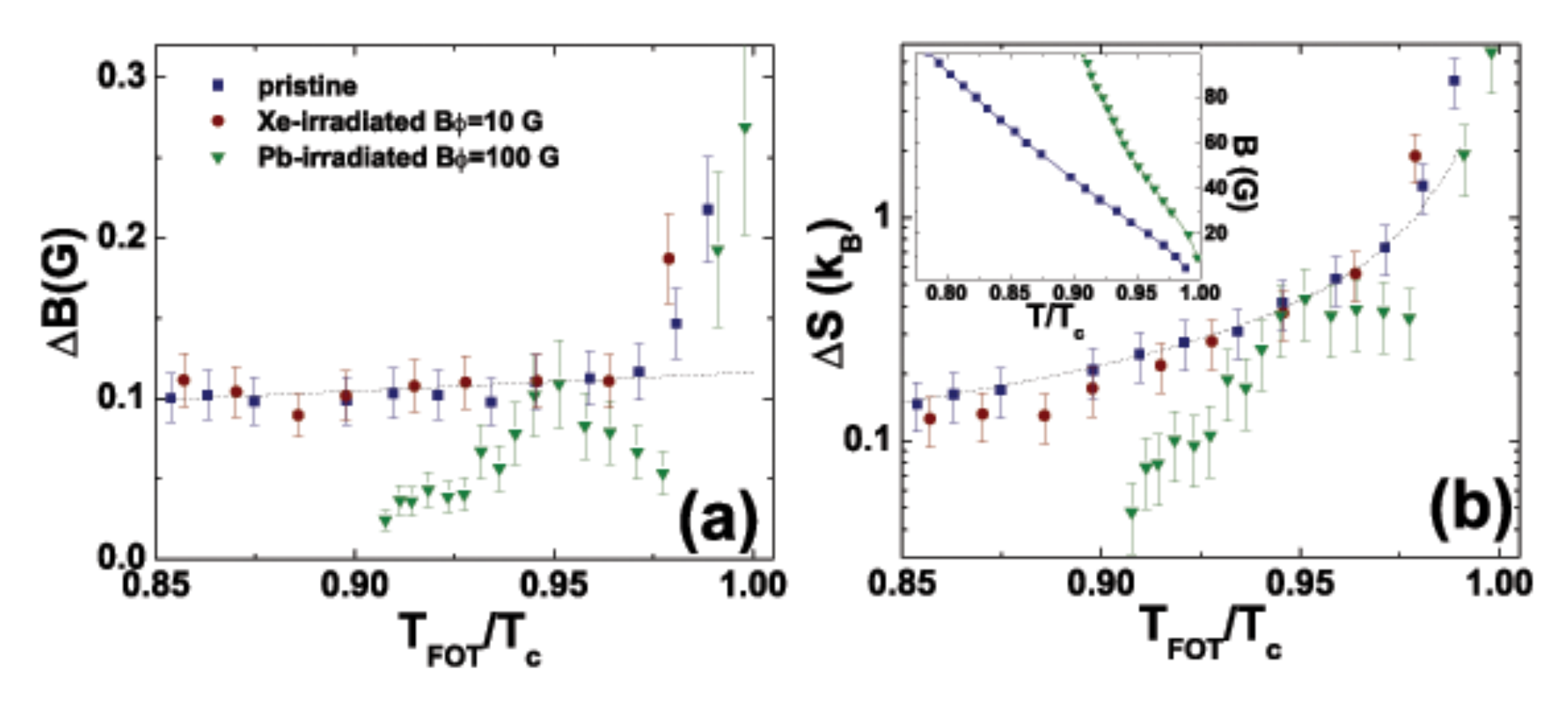}
\caption{Reduced-temperature evolution of the (a) enthalpy
$\propto \Delta B$ and (b) entropy $\Delta S$ at the first-order
vortex solidification transition for pristine as well as Xe and
Pb-irradiated samples. The dotted lines are fits to the data with
a theoretical model that only takes into account electromagnetic
coupling between pancake vortices at the transition (see text).
Insert: vortex phase diagram of the samples. (color figure online). \label{figure3}}
\end{figure*}

Figure\,\ref{figure2} (a) shows, in accordance with the
literature, that introducing a small amount of quenched disorder
by means of heavy-ion irradiation spans the solid vortex phase in
a larger temperature region. This is observed in the
$T'$ measurements showing a shifting of the
paramagnetic peak towards higher $T/T_{\rm c}$ in the case of
Xe-irradiated $B_{\Phi}=10$\,Gauss with respect to pristine
samples. Figure \,\ref{figure2} (b) shows one of the main results
of this work: The FOT persists for a larger amount of quenched
disorder corresponding to $B_{\Phi}=100$\,Gauss, in contrast to
magneto-optical results \cite{Banerjee2003,Menghini2003}. The
paramagnetic peaks in the Pb-irradiated $B_{\Phi}=100$ sample are
detected up to more than 100\,Gauss and shift towards higher
$T/T_{\rm c}$ in comparison to the pristine sample for a
fixed applied field. For instance, in this sample, the solid
vortex phase spans a $T/T_{\rm c}$ region 20\% larger than for the
pristine sample at 100\,Gauss (see insert to Fig.\,\ref{figure3}
(b)). The discrepancy between our data and previous one has
its origin in the local induction resolution in both experiments,
being not better than 1\,Gauss for magneto-optics data and of
the order of 5\,mGauss in our AC magnetometry measurements.

The jump in $B$ entailed in the FOT can be obtained from AC
measurements taking into account that for $T \gtrsim T_{\rm FOT}$
the transmittivity $T' \sim B'/H_{\rm ac}$. Following
thermodynamic considerations as in Ref.\,\cite{Morozov1996}, and
considering that the magnitude of $H_{\rm ac}$ is small, the
transmittivity can be approximated by $T'(T_{\rm FOT})= 1+\frac{2
\Delta B}{\pi H_{\rm ac}}$. Inverting this equation allows the
determination of $\Delta B$ as a function of $T_{\rm FOT}/T_{\rm c}$ as
shown in Fig.\,\ref{figure3} (a). It is important to point
out that the value of $\Delta B$ obtained in this way is equal to
the obtained in DC magnetization loops \cite{Morozov1996,Dolz2014}. In the case of the pristine and
$B_{\Phi}=10$\,Gauss samples $\Delta B$ increases roughly linearly
up to $T_{FOT} \sim  T_{c}$ within the error bars. For the
$B_{\Phi}=100$\,Gauss sample the data do not follow this linear
evolution, particularly at low temperatures. A theoretical work
\cite{Dodgson1998} predicted a linear evolution of $\Delta B$  by
considering that the inter-layer coupling is only dominated by
electromagnetic interactions between pancake vortices
undergoing large thermal fluctuations in a material with
negligible pinning. This work proposed that

$$\Delta B= \mu ' \frac{k_{\rm B} T_{\rm FOT}}{\Phi_{\rm 0} d},$$

\noindent where $k_{\rm B}$ is the Boltzmann constant, $\Phi_{\rm
0}$ the flux quantum, $d \approx 15 \AA$ the distance between CuO
planes and $\mu '$ a numerical constant. The
results in the pristine and $B_{\Phi}=10$\,Gauss samples are
reasonably well fitted with this functionality, in contrast to the case
of the $B_{\Phi}=100$\,Gauss sample.

A non-linear evolution of $\Delta B$ was previously reported for a
pristine Bi$_{2}$Sr$_{2}$CaCu$_{2}$O$_{8}$ sample in
Ref.\,\cite{Zeldov95a}, with this magnitude decreasing to a
quarter of its low-temperature value for $T_{\rm FOT}/T_{\rm c}
\geq 0.93$. The mentioned theoretical work interpreted this
as an indication that electromagnetic interactions alone
cannot account for the location of the FOT line close to $T_{\rm
c}$ and therefore suggested a crossover to a Josephson-coupling
regime \cite{Dodgson1998}. Our $\Delta B$ data depart from the
linear behavior at low temperatures when the magnitude of quenched disorder is
increased. Since the theoretical proposal does not take into
account the effect of pinning, our results in pristine and
irradiated samples challenge the existence of such a crossover.

The entropy-jump at $T_{\rm FOT}$ obtained from the
Clausius-Clapeyron relation is

$$\Delta S= -\frac{\phi_{\rm 0} d}{4\pi}\frac{\Delta B}{B_{\rm
FOT}} \, \frac{dH_{\rm FOT}}{dT} .$$

\noindent Figure \ref{figure3} (b) shows that the $\Delta S$
entailed in the vortex solidification FOT diverges for $T \sim
T_{\rm c}$ for pristine as well as for irradiated
Bi$_{2}$Sr$_{2}$CaCu$_{2}$O$_{8}$ samples. Applying the same
thermodynamic relation to obtain the entropy of the
theoretical system that only takes into account electromagnetic
coupling holds
$$\Delta S= \frac{\mu}{\pi} \frac{k_{\rm
B}}{[1-(T_{\rm FOT}/T_{\rm c})^2]}.$$

\noindent The fit of our experimental data with this function is
shown with the dotted line of Fig.\,\ref{figure3}. Note that this
function has only one fitting parameter, the multiplicative
constant $\mu$. The  $\Delta S$ values for the pristine sample
follow this theoretical functionality but apart from it on
increasing the magnitude of quenched disorder. The fact that this
departure between the experimental data and the theoretical model
happens at low temperatures is also at odds with ascribing it to a
crossover to Josephson-dominated coupling.

\section{Conclusions}

Our main result of the persistence of the FOT for
$B_{\Phi}=100$\,Gauss calls for a revision of the nature of the
vortex solidification transition in the case of extremely-layered
superconducting samples with a moderate magnitude of quenched
disorder. Our findings do not challenge the accepted picture of a
change from first to second-order transition through a critical
point for high-fields or alternatively high-magnitude of quenched
disorder. However we put in evidence that the fields and
$B_{\Phi}$ for which these critical points occur have to be
re-examined by means of local magnetic techniques with improved
resolution. In addition, the departure of the enthalpy of the
transition from a linear behavior in $T_{\rm FOT}/T_{\rm c}$ when
increasing the magnitude of quenched disorder suggests that it is not
straightforward to ascribe this phenomenology to a crossover
between an electromagnetic-to-Josephson coupling of pancake
vortices. A  theoretical description taking also into account the
effect of pinning is therefore required in order to properly
describe the evolution of the thermodynamic magnitudes in the
first-order vortex solidification transition.

\end{document}